      [1999/12/01 v1.4c Il Nuovo Cimento]
\documentclass{cimento}

\usepackage{graphicx,epsf,epsfig}  % got figures? uncomment this
\title{Search for New Physics the Fermilab Tevatron $p\bar p$~ Collider }
\author{S.~Rolli\from{ins:x}}
\instlist{\inst{ins:x} Tufts University,\\
4 Colby St, Medford, MA 02155, USA, \\
currently at:\\
U.S. Department of Energy, Office of High Energy Physics, \\
Washington, DC 20585, USA}
\PACSes{\PACSit{14.80.Bn}{}}
\begin{document}

\maketitle

\begin{abstract}
We report on selected recent results from the CDF and D0 experiments
on searches for physics beyond the Standard Model using data from the
Tevatron collider running $p\bar p$ collisions at $\sqrt{s}$ = 1960 GeV.

\end{abstract}

\section{Introduction}

Over the past decades the Standard Model (SM) of particle physics has been surprisingly
successful.
Although the precision of experimental tests improved by orders
of magnitude no significant deviation from the SM predictions has
been observed so far.
Still, there are many questions that the Standard Model does not
answer and problems it can not solve.
Among the most important ones are the origin of the electro-weak
symmetry breaking, hierarchy of scales, unification of fundamental forces
and the nature of gravity.
Recent cosmological observations indicates that the SM particles only account for ~4\% of the matter of the
Universe.
Many extensions of the SM (Beyond the Standard Model, BSM) 
have been proposed to make the theory more complete and solve some of the
above puzzles. Some of these extension includes SuperSymmetry (SUSY), Grand Unification Theory (GUT)
and Extra Dimensions. 
At CDF and D0 we search for evidence of such processes in proton-antiproton collisions
at $\sqrt(s)$ = 1960 GeV. 
The phenomenology of these models is very rich, although the cross sections
for most of these exotic processes is often very small compared to those of SM processes at hadron colliders.
\begin{figure}[ht]
\begin{center}
\includegraphics[height=1.2in,width=2.0in]{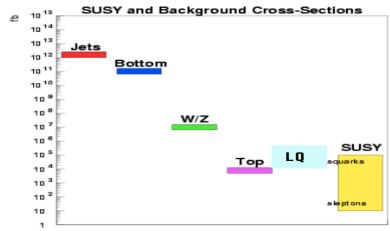}
\caption{Cross sections for typical SM processes at the TeVatron and exotic physics}
\end{center}
\end{figure}
It is then necessary to devise analysis strategies that would allow to disentangle the small interesting signals,
often buried under heavy instrumental and/or physics background. 
Two main approaches to search for physics beyond the Standard Model are used 
in a complementary fashion: model-based analyses and signature based studies.
In the more traditional model-driven approach, one picks a favorite 
theoretical model and/or a process, and the best signature is chosen. 
The selection cuts are optimized based on acceptance studies performed using simulated signal events. 
The expected background is calculated from data and/or Monte Carlo and, 
based on the number of events observed in the data, a discovery is made or 
the best limit on the new signal is set.
In a signature-based approach a specific signature is picked (i.e. dileptons+X) and the 
data sample is defined in terms of known SM processes. A signal region (blind box) 
might be defined with 
cuts which are kept as loose as possible and the background predictions in the signal 
region are often extrapolated
from control regions.  
Inconsistencies with the SM predictions will provide indication of possible new physics. 
As the cuts and acceptances are often calculated independently from a model, 
different models can be tested against the data sample. It should be noticed that
the comparison with a specific model implies calculating optimized acceptances for a specific BSM signal.
In signature-based searches, there is no such an optimization. 
Both the experiments have followed a somehow natural approach in pursuing analysis looking at final state signatures 
characterized by relatively simple physics objects (for example lepton-only final state, where the selection of the leptons
is straightforward and can be easily checked with the measurement of electroweak boson production cross sections)
and proceeding onto more complex final state, including jets and heavy flavor. Here more sophisticated identification
techniques need to be used and issues like jet energy scale calibration play an important role in determining the final
result.
Given the limited space available for this proceeding, we will focus here on  
few selected results. Further results are described in~http://ncdf70.fnal.gov:8001/presentations/LaThuile2011\_Rolli.pdf.

\section{Search for new physics in dileptons final states}
This is a typical example of a signature-based search for new physics.
Final states consisting of dileptons are a straightforward signature where to look for new physics, as several resonant 
states can appear as enhancement of the Drell-Yan cross section. 
The analysis strategy is very simple: the invarian mass distribution of
the dilepton system is compared to the SM expectations, as shown in figure~2 and 3.
Only identification cuts to select a pair of high $P_T$ leptons are placed.

Both CDF\cite{ref:cdf-dielectron-paper,ref:cdf-dimuons-paper} and D0\cite{ref:d0-dielectron-paper} have been studying
the dilepton invariant mass distribution.
The most recent result is a search for new dielectron mass resonances 
using 5.7 fb$^{-1}$ of data recorded by the CDF II detector. 
No significant excess over the expected standard model prediction is observed. In
this dataset, an event with the highest dielectron mass ever observed (960 GeV/c2) has been recorded. 
The results are intepreted in the framework of the Randall-Sundrum (RS) model\cite{ref:rs-graviton-paper}. 
Combined with a similar search performed with 5.4-fb$^{-1}$ of diphoton data \cite{ref:cdf-diphoton-paper} 
the RS-graviton mass limit for the coupling k/M P$_l$ = 0.1 is 1058 GeV/c$^2$ at 95\% CL,
making it the strongest limit to date.
A similar search is performed in the dimuon channel using 4.3 fb$^{-1}$ of data and no excess is observed. The result
is interpreted in terms of Z' production and limits are set on several Z' production scenario: such limits are 
extending to the kinematical reach of the Tevatron ( sequential SM Z' limit is set for example to 1071 GeV/c$^2$ 
at 95\% CL, making it
one of the most stringent in this channel).

\begin{figure}
\begin{center}
  \begin{minipage}{2.5in}
\psfig{figure=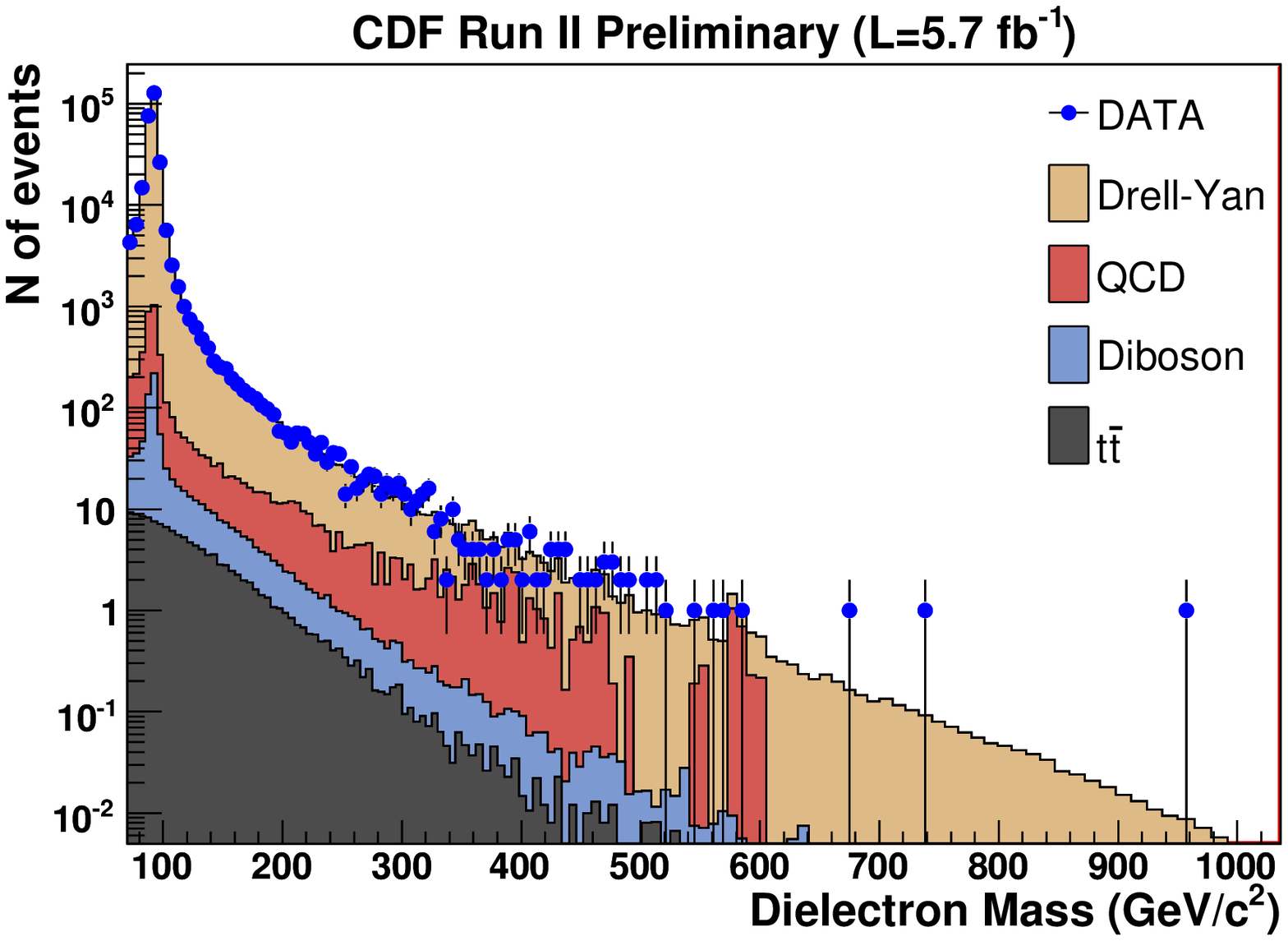,height=2.0in}
\caption{ {\scriptsize Inclusive dielectron mass spectrum at CDF 
}}
    \label{figure1}
  \end{minipage}
\hfill
  \begin{minipage}{2.5in}
\psfig{figure=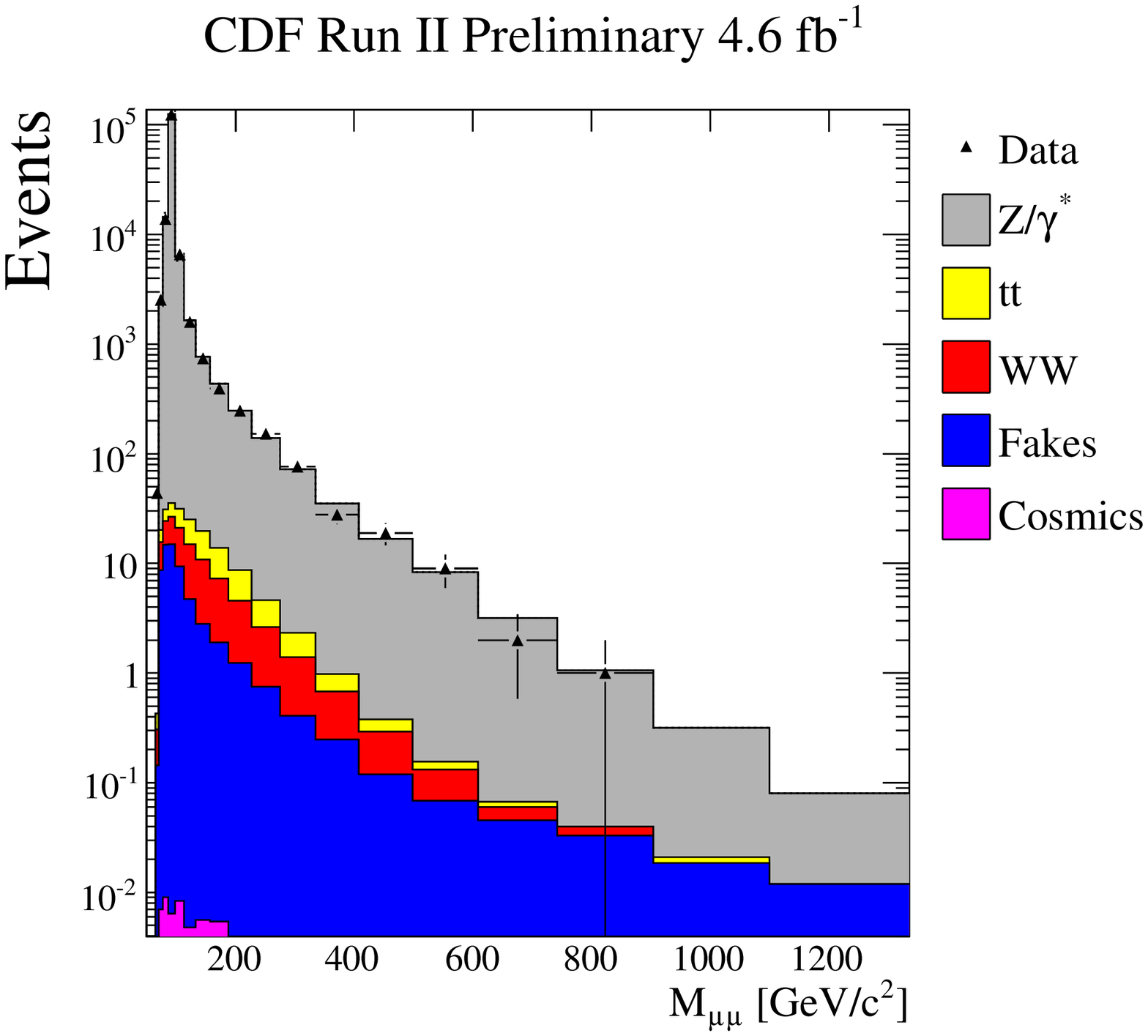,height=2.0in}
\caption{{\scriptsize Inclusive dimuons mass spectrum at CDF 
}}
    \label{figure2} 
  \end{minipage}
\hfill
\end{center}
\end{figure}

\section{Search for extra vector bosons and diboson resonances}
A recent result by the  D0 collaboration\cite{ref:d0-dibosons-paper} concerns the 
search for resonant WW or WZ production. The dataset used corresponds to 5.4 fb$^{-1}$ of integrated luminosity 
collected by the D0 experiment. The search for these resonances in 
the diboson decay channel covers the possibility that their coupling to leptons 
may be lower than the value predicted by the SM.
The data are consistent with the standard model background expectation, and limits are set 
on a resonance mass using the sequential standard model (SSM) W′ boson and the Randall-Sundrum model graviton G 
as benchmarks. D0 excludes a SSM W' boson in the mass range 180-690 GeV and a Randall-Sundrum graviton 
in the range 300-754 GeV at 95\% CL.
There are two recent direct searches for
WZ or WW resonances by	the CDF	and D0	collaborations\cite{ref:d0-olddiboson-paper,ref:cdf-diboson-paper}
that exclude WZ resonances with mass
below 516 and 520 GeV, respectively, and an RS graviton	$G\to W W$ resonance with mass less than
607 GeV. 
Indirect searches for new physics in the WW and	WZ diboson systems through measurements
of the triple gauge couplings also show no deviation  
from the SM predictions\cite{ref:d0-indirectWZ-paper1,ref:d0-indirectWZ-paper2,ref:d0-indirectWZ-paper3}
Finally the CDF collaboration has very recently excluded M(W') $<$ 1.1 TeV, when assuming the W' boson decays as in the 
SM\cite{ref:cdf-wprime-paper}.

\begin{figure}
\begin{center}
  \begin{minipage}{2.5in}
\psfig{figure=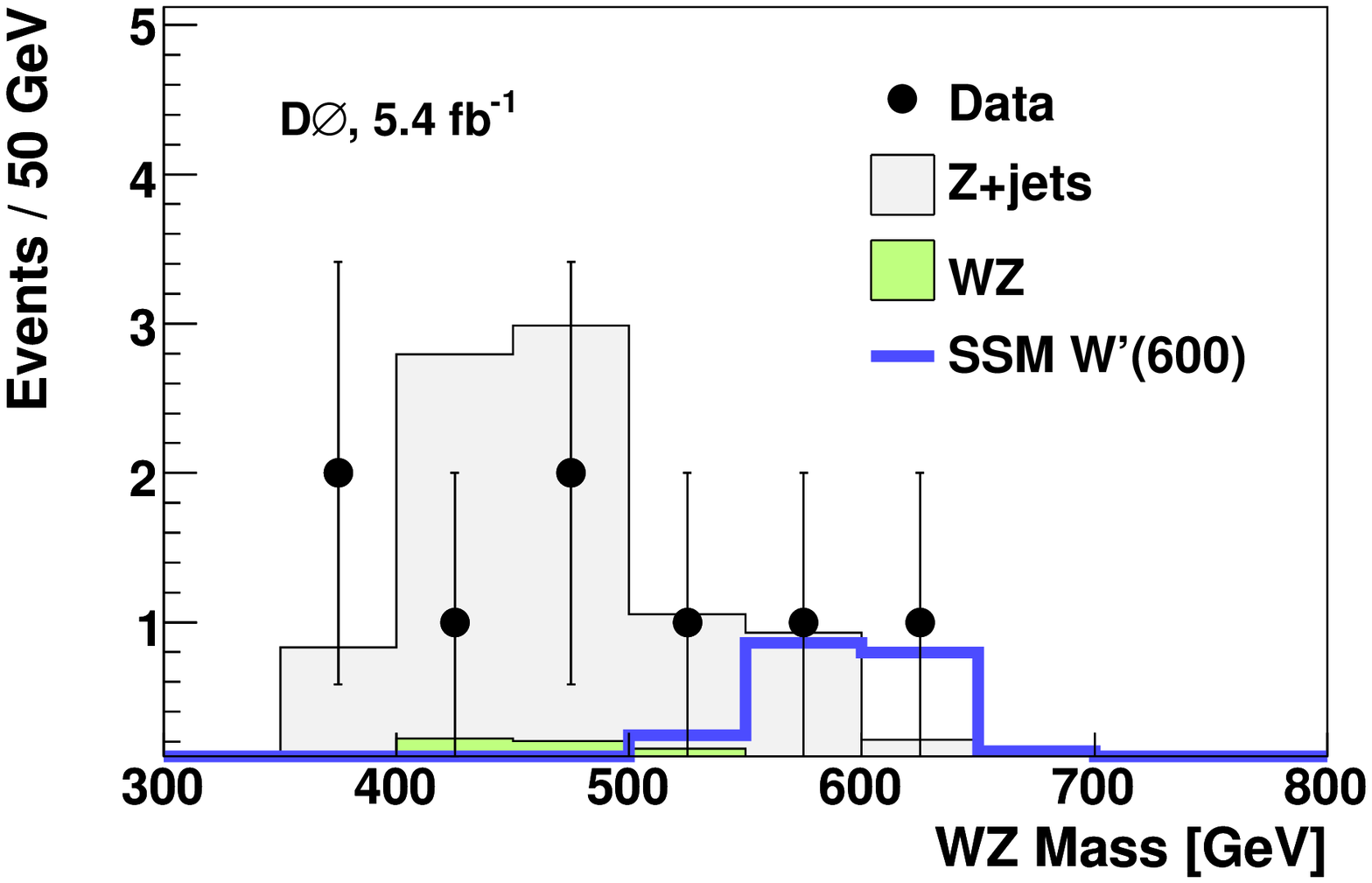,height=2.0in}
\caption{ {\scriptsize Reconstructed WZ mass in the  l$\nu$jj, lljj, lll$\nu$ channels, D0 collaboration.
}}
    \label{figure3}
  \end{minipage}
\hfill
  \begin{minipage}{2.5in}
\psfig{figure=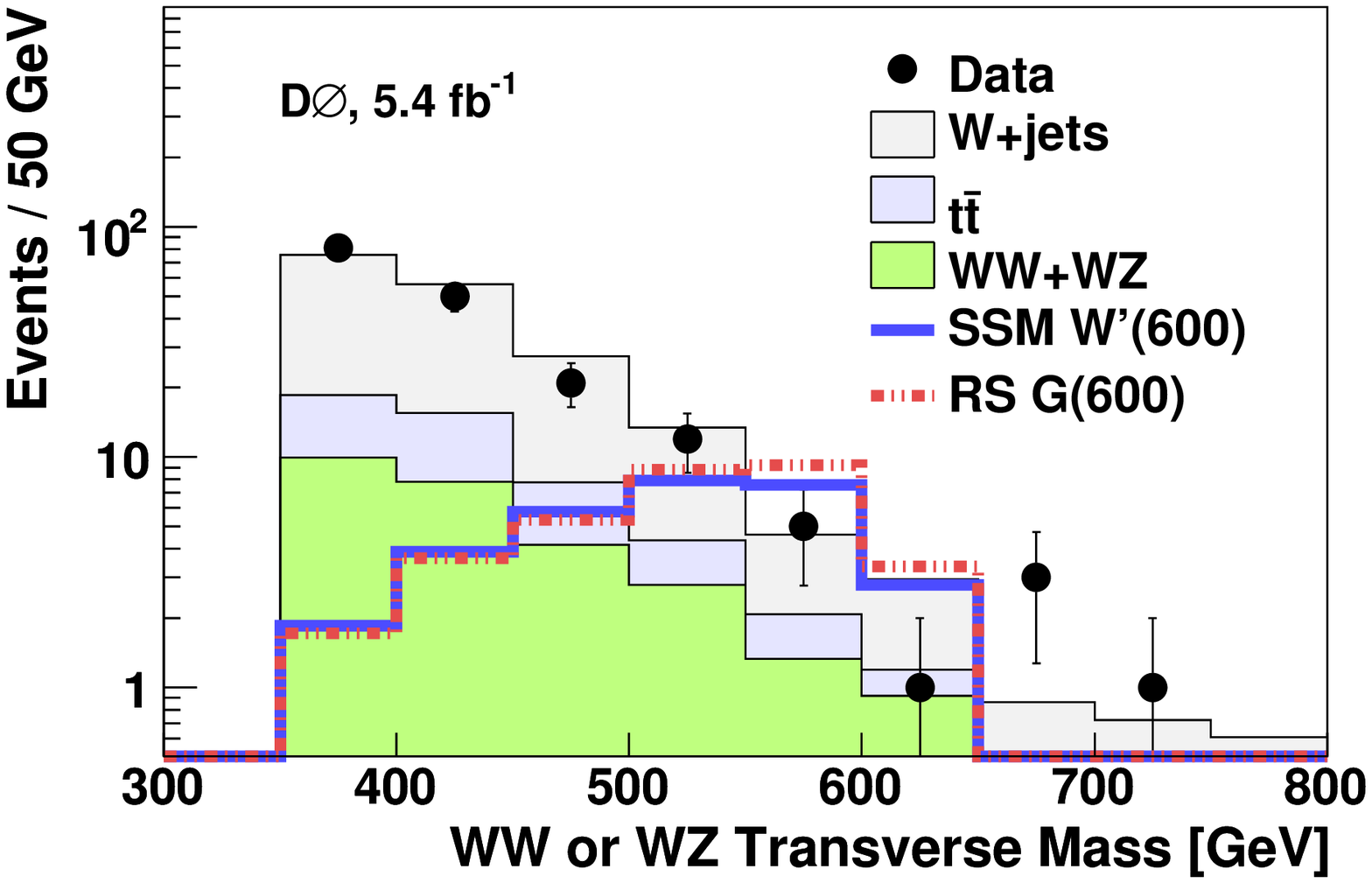,height=2.0in}
\caption{{\scriptsize Reconstructed WZ or WW transverse mass in the  l$\nu$jj channel, D0 collaboration.
}}
    \label{figure4} 
  \end{minipage}
\hfill
\end{center}
\end{figure}

\section{Search for new physics in complex final states}

\subsection{gamma plus jets}

Many new physics models predict mechanisms that could produce a $\gamma$+jets signature. 
CDF searches in the $\gamma$+jets channel, independently of any model, for new physics using 
4.8 $fb^{-1}$ of CDF Run II data\cite{ref:cdf-gammajet-paper}. 
A variety of techniques are applied to estimate the Standard Model expectation and non-collision backgrounds. 
Several kinematic distributions are examined, including photon ET , invariant masses, and total transverse energy 
in the event for discrepancies with predictions from the Standard Model. The data are found to be consistent with 
Standard Model expectations. This global search for new physics in $\gamma$+jets channel reveals no significant 
indication of physics beyond Standard Model.

\begin{figure}
\begin{center}
  \begin{minipage}{2.5in}
\psfig{figure=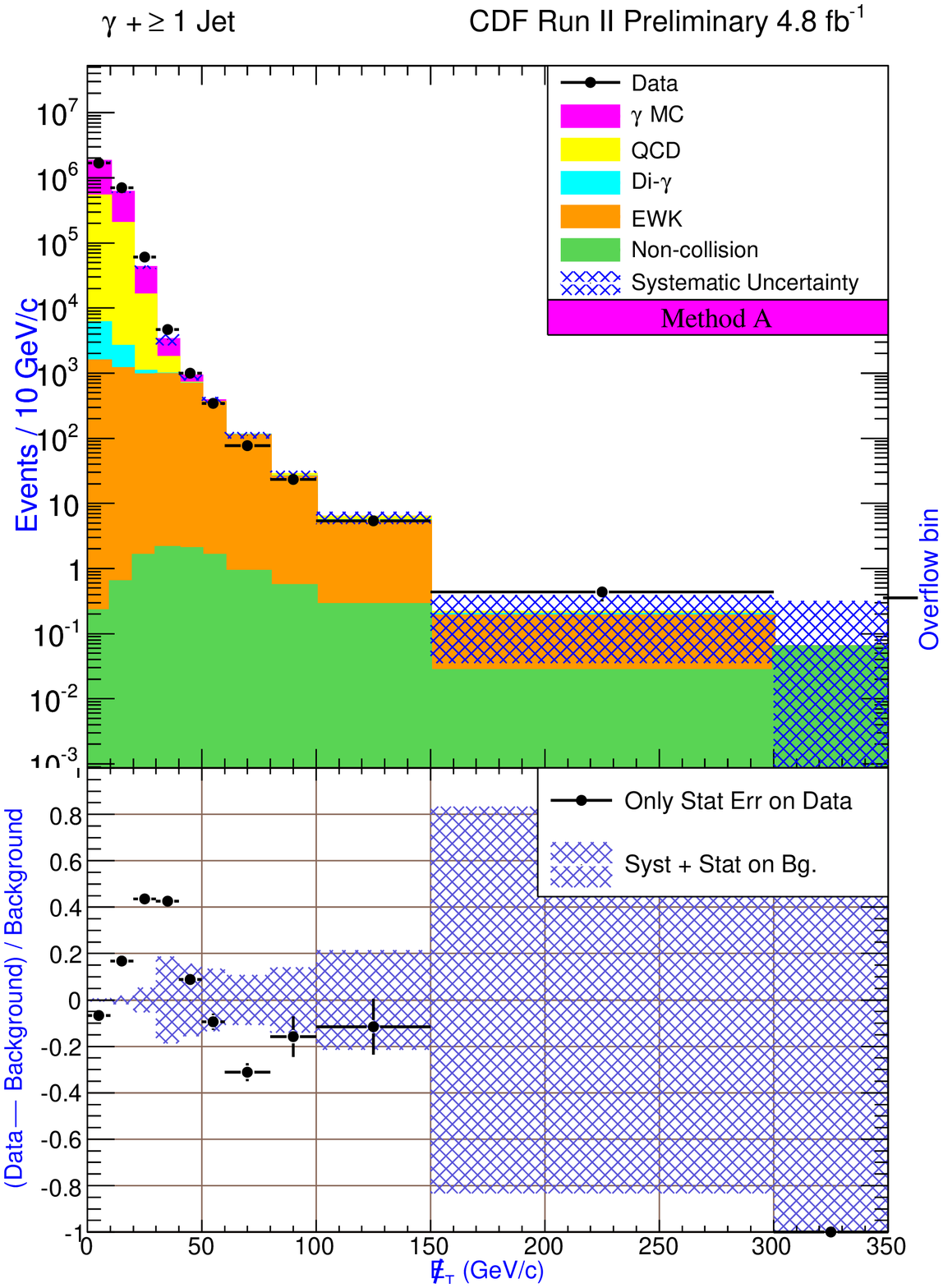,height=2.0in}
\caption{ {\scriptsize Missing Energy distribution for $\gamma \pm $ 1 jet, CDF collaboration.
}}
    \label{figure5}
  \end{minipage}
\hfill
  \begin{minipage}{2.5in}
\psfig{figure=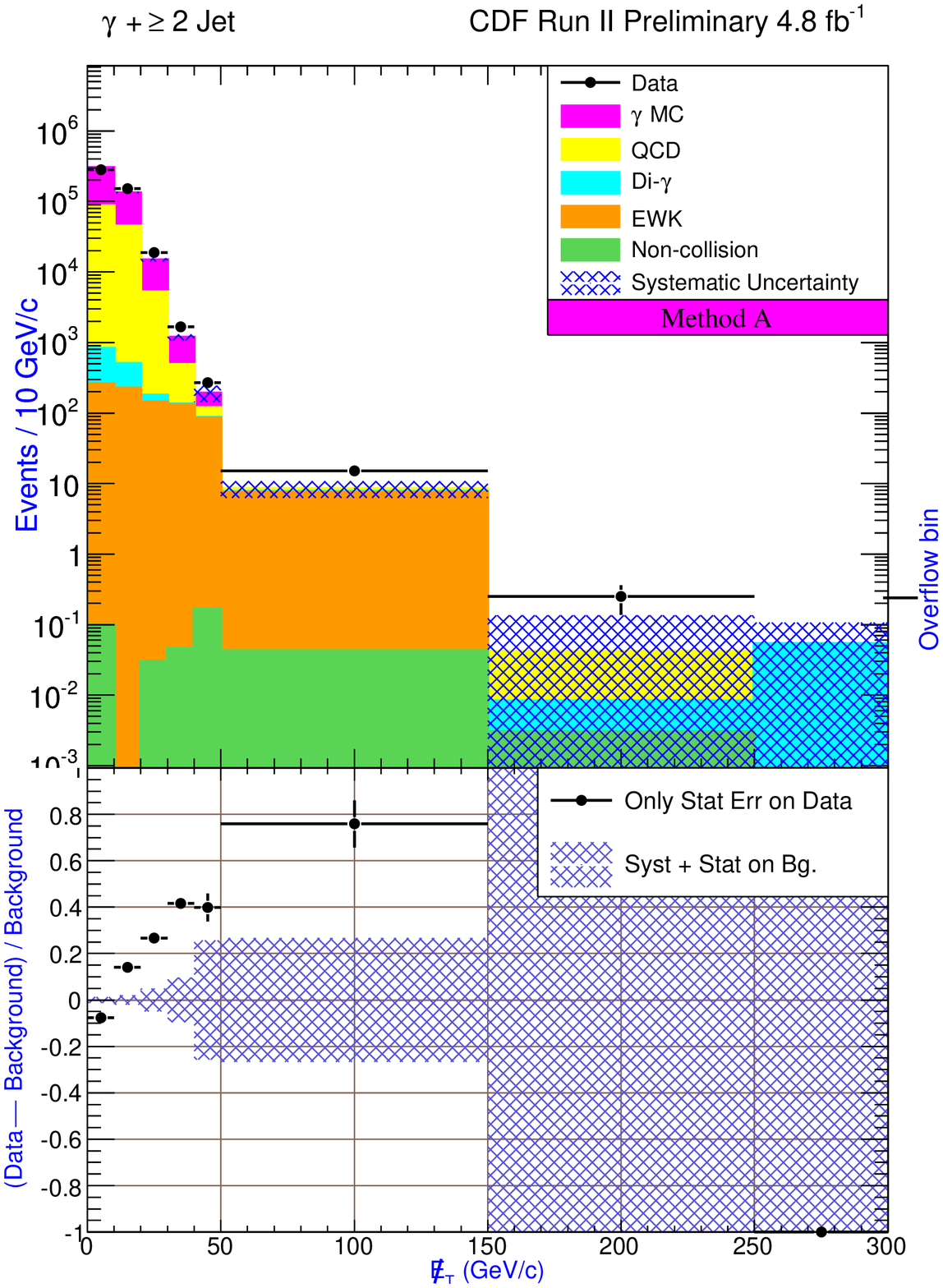,height=2.0in}
\caption{ {\scriptsize Missing Energy distribution for $\gamma \pm $ 2 jet, CDF collaboration.
}}
    \label{figure6}
  \end{minipage}
\hfill

\end{center}
\end{figure}

\subsection{gamma plus b-jets plus MET + leptons}

A search for anomalous production of the signature l+$\gamma$+b-quark+MET (l$\gamma$ MET b) 
has been performed by using 6.0 fb$^{-1}$ of data taken with the CDF detector\cite{ref:cdf-ttgamma-paper}. 
In addition to the l$\gamma$ MET b signature-based search, CDF also presents for the first time a search for 
top pair production with an additional radiated photon, $t \bar t + \gamma$. 
85 events of l$\gamma$ MET b versus an expectation of 99.1 $\pm$ 7.61 events. 
Additionally requiring the events to contain at least 3 jets and to have a total transverse energy of 200 GeV, 
CDF observes 30 $t \bar t  \gamma$ candidate events versus an expectation from non-top standard model (SM) sources 
of 13.0 $\pm$ 2.1. Assuming the difference between the observed number and the predicted non-$t \bar t  \gamma$ SM total 
is due to $t \bar t  \gamma$ production, the collaboration measures the $t \bar t  \gamma$ cross section 
to be 0.18 ± 0.07 (stat.)$\pm$ 0.04 (sys.)$\pm$ 0.01 (lum.) pb. 
We also measure a ratio of the $t \bar t  \gamma$ cross section to the $t \bar t $ cross section to be 
0.024 $\pm$ 0.009.

 \begin{figure}[h]
\begin{center}
%  \begin{minipage}{2.5in}
\psfig{figure=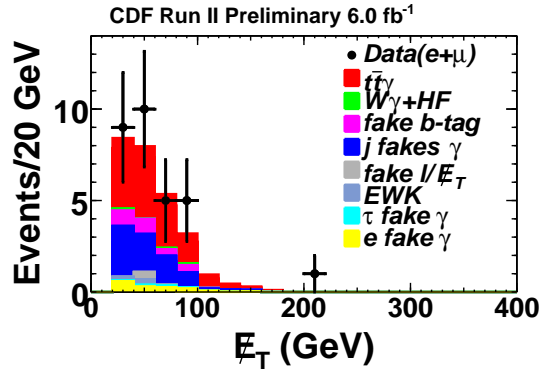,height=2.0in}
 \caption{ {\scriptsize Missing Energy distribution in l $\gamma$ MET b events, CDF collaboration.
 }}
    \label{figure7}
%    \end{minipage}
\end{center}
\end{figure}

\subsection{Multijets resonances}

A new analysis from CDF has been performed to search for 3-jet hadronic 
resonances in 3.2 fb$^{-1}$ of data\cite{ref:cdf-multijets-paper}. 
Typical searches for new physics require either leptons and/or missing transverse energy, 
however, they might be blind to new physics which have strong couplings and therefore decay into quarks and gluons. 
The CDF collaboration used  3.2 fb$^{-1}$ of data in a model-independent search that reconstructs hadronic resonances 
in multijet final states. Although the analysis is not optimized for a specific model of new physics,
we use as a possible benchmark, R-parity violating supersymmetric 
(RPV SUSY) gluino pairs production, with each gluino decaying into three objects. 
Since no significant excess is observed 
in the data a 95\%C.L.limit is set on 
$\sigma(p \bar p \to XX ) \times Br(\tilde g \tilde g \to 3 jets + 3 jets)$  where X = $\tilde g, \tilde q$, as a 
function of the  gluino invariant mass.
To extract signal from the multijet QCD background, kinematic quantities and correlations are used
to create an ensemble of jet combinations. Incidentally, the all-hadronic $t \bar t$ decay has a signature 
similar to the signal searched for in this analysis. The biggest challenge of the analysis is the large 
QCD background that 
accompanies multijet resonances. A data driven approach is used to parameterize such background.
An ensemble consists of 20 (or more) possible jet triplets from the  $\ge$ 6 hardest jets in the event. 
For every event, we calculate each jet triplet invariant mass, $M_{jjj}$ , and scalar sum $p_T$ , 
$\Sigma_{jjj} |pT |$. Using the distribution of $M_{jjj}$  vs. $\Sigma{jjj} |pT |$ ensures that the correct combination of jets in
pre-defined kinematic regimes is reconstructed, since the 
incorrect (uncorrelated) triplets tend to have $M_{jjj}$ = $\Sigma{jjj} |pT |$. The correct ( correlated) triplet produces
a horizontal branch in the signal at approximately the invariant mass of the signal that is not present for the background
as can be seen in figures 9,10,11,12.

 \begin{figure}[h]
\begin{center}
  \begin{minipage}{2.5in}
\psfig{figure=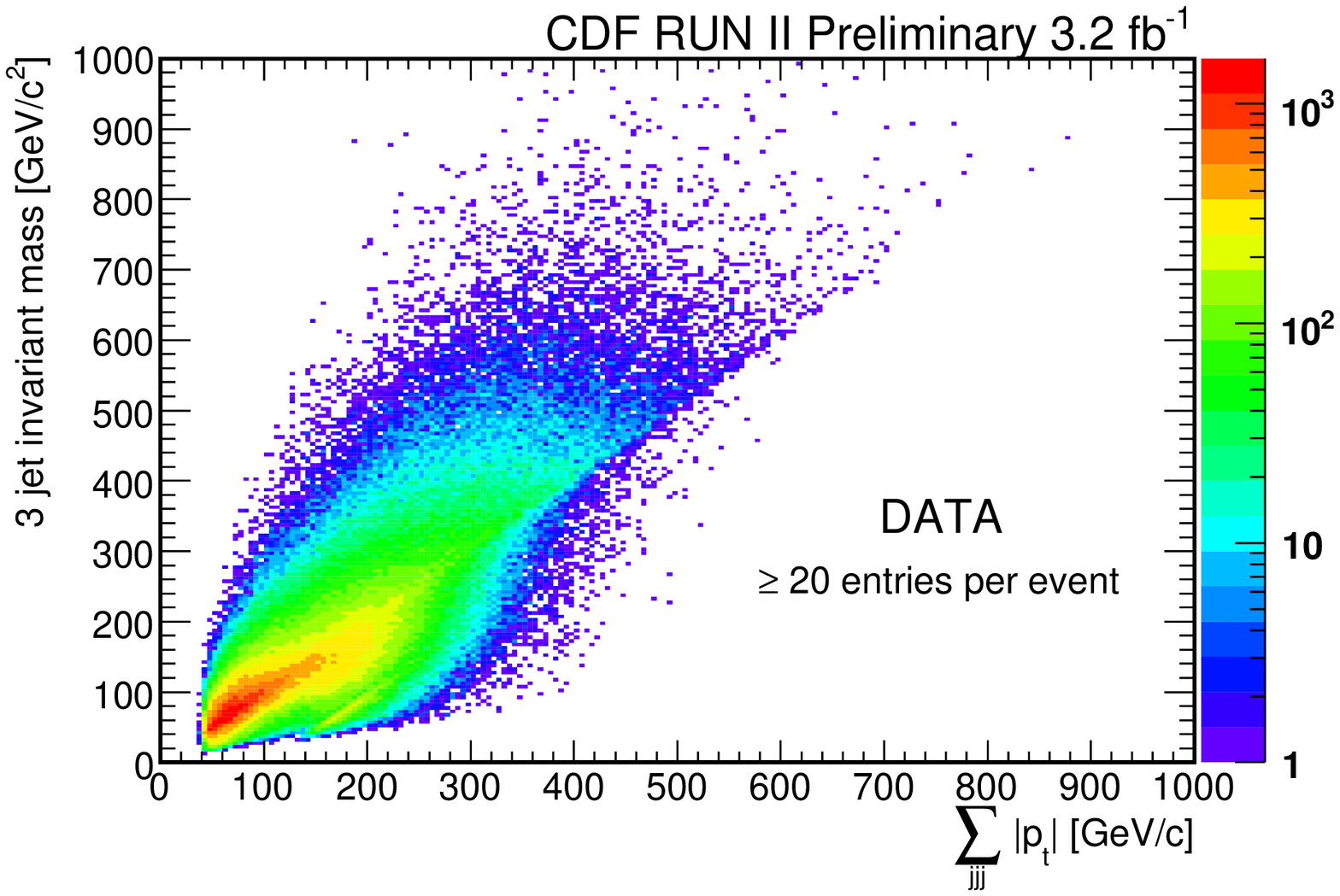,height=2.0in}
 \caption{ {\scriptsize Distributions of $M_{jjj}$ versus $\Sigma_{jjj}|pT|$ multiple entry($\ge 20$).data, 
}}
    \label{fig:multijets1}
    \end{minipage}
\hfill
  \begin{minipage}{2.5in}
\psfig{figure=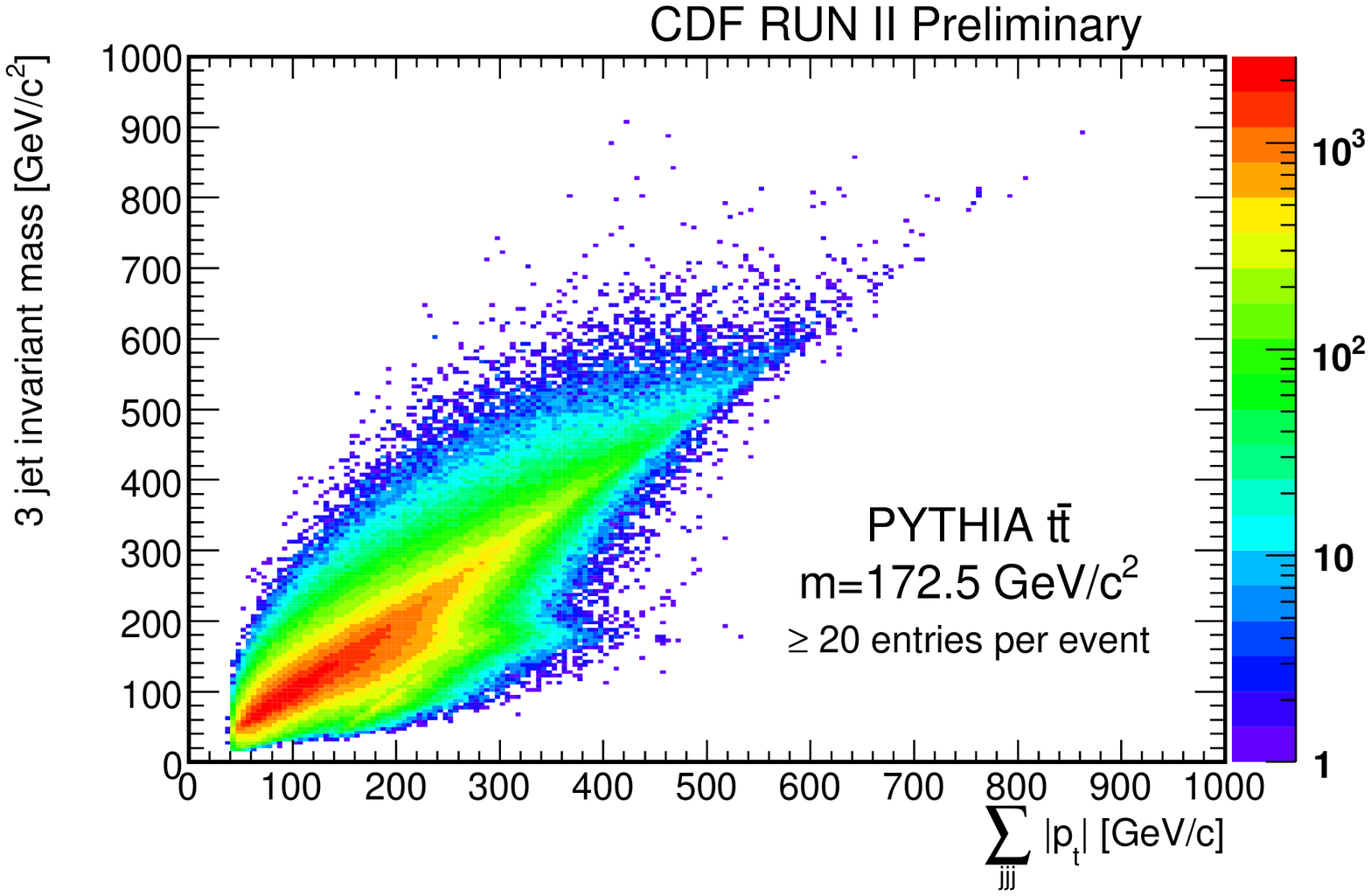,height=2.0in}
 \caption{ {\scriptsize Distributions of $M_{jjj}$ versus $\Sigma_{jjj}|pT|$ multiple entry($\ge 20$).pythia $t \bar t$ (m=172.5GeV/c$^2$), 
}}
    \label{fig:multijets2}
     \end{minipage}
%\end{center}
%\end{figure}
\vfill
% \begin{figure}[h]
%\begin{center}
  \begin{minipage}{2.5in}
\psfig{figure=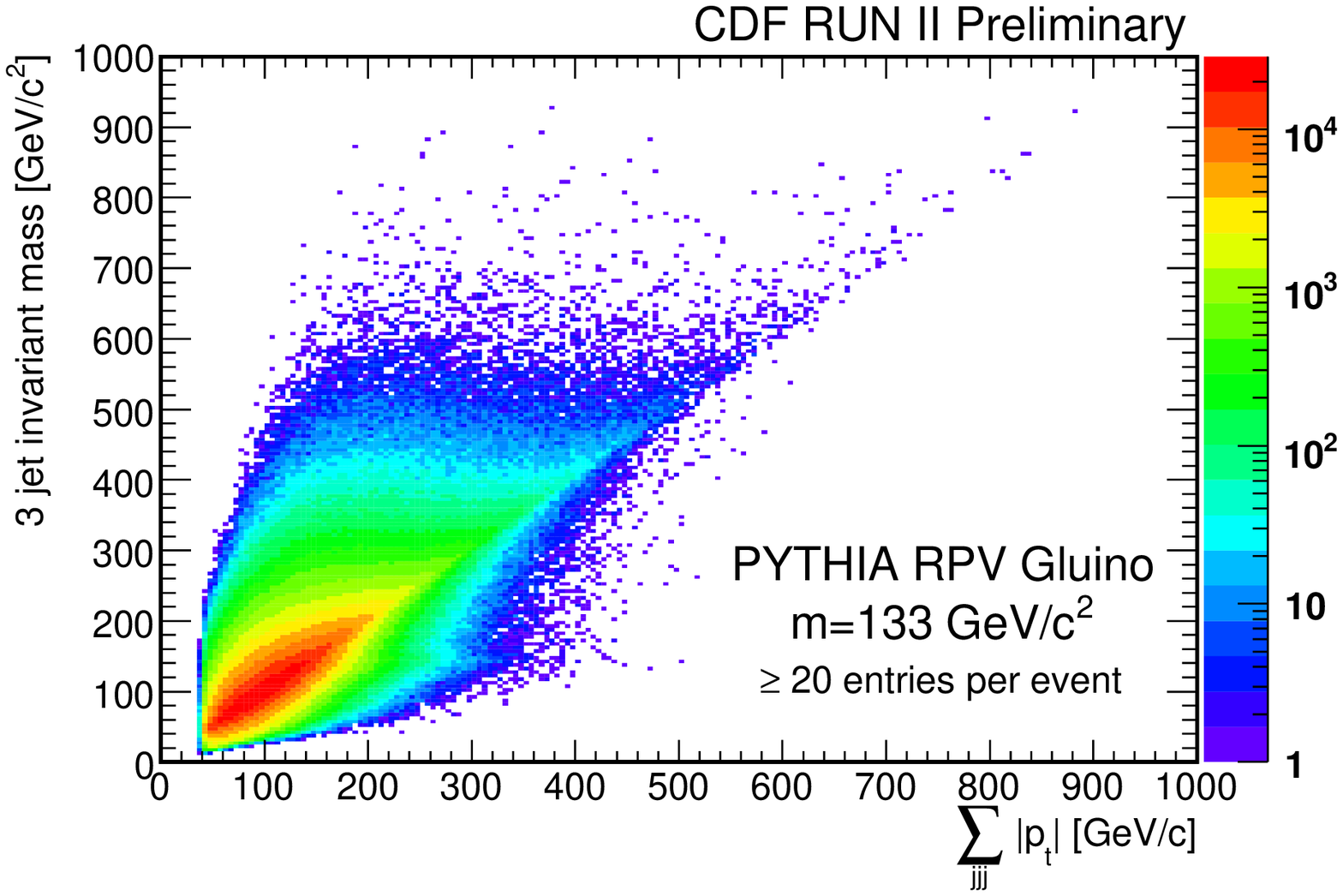,height=2.0in}
 \caption{ {\scriptsize Distributions of $M_{jjj}$ versus $\Sigma_{jjj}|pT|$ multiple entry($\ge 20$).pythia RPV gluino ( m= 133 GeV/$c^2$),
 }}
    \label{fig:multijets3}
    \end{minipage}
\hfill
  \begin{minipage}{2.5in}
\psfig{figure=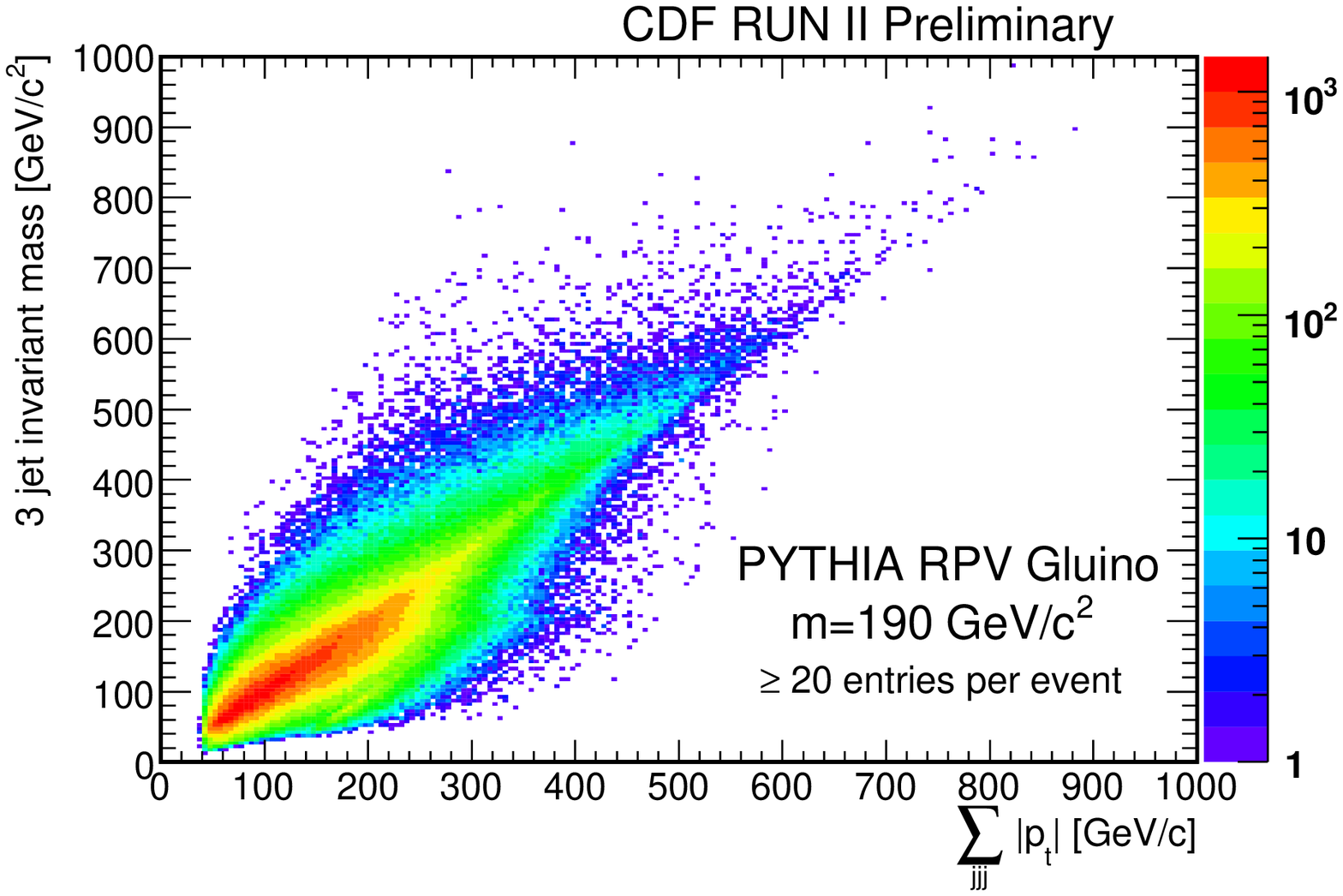,height=2.0in}
 \caption{ {\scriptsize Distributions of $M_{jjj}$ versus $\Sigma_{jjj}|pT|$ multiple entry($\ge 20$).pythia RPV gluino ( m= 190 GeV/$c^2$)
}}
    \label{fig:multijets4}
     \end{minipage}
\end{center}
\end{figure}

 \begin{figure}[h]
\begin{center}
%  \begin{minipage}{2.5in}
\psfig{figure=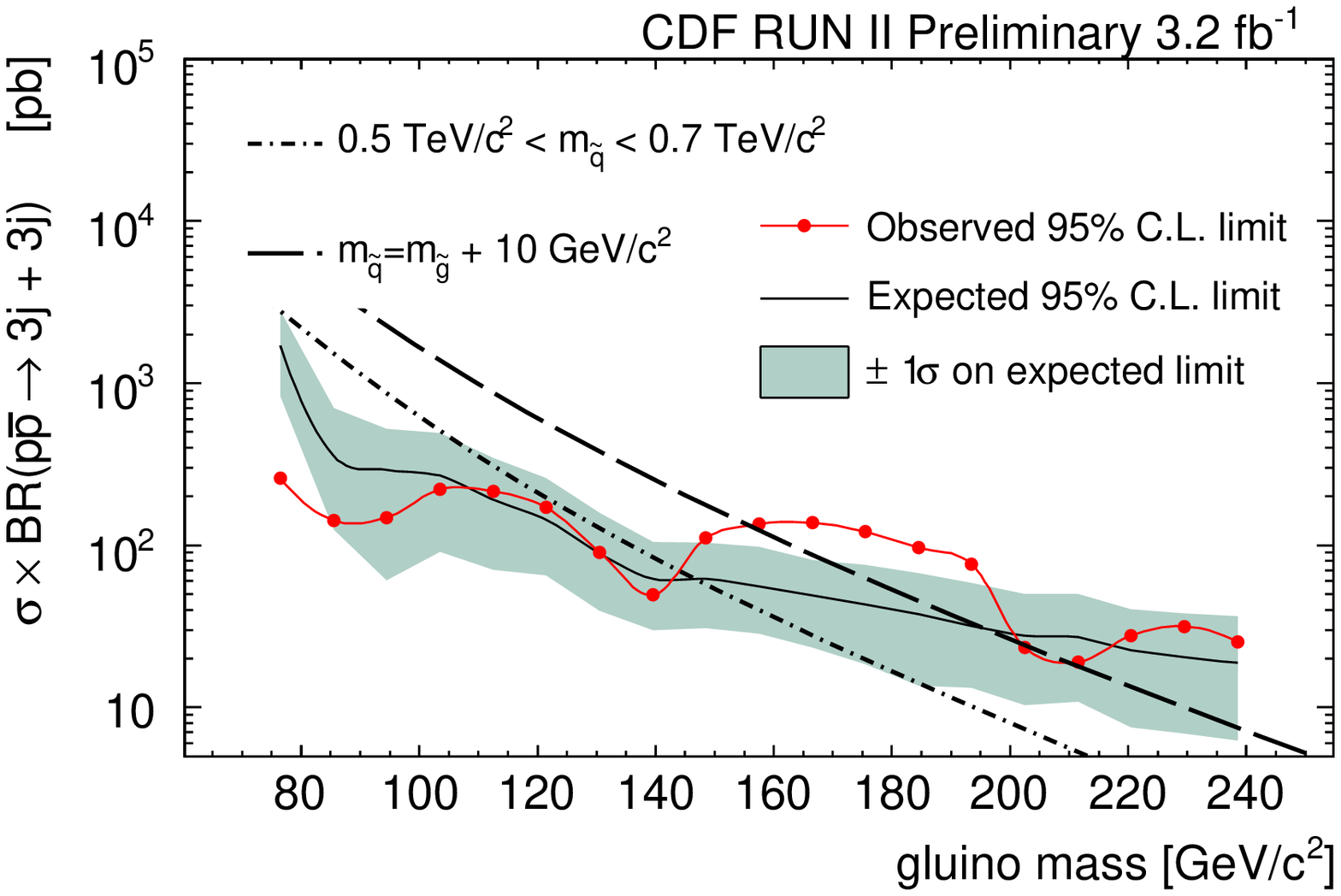,height=2.0in}
 \caption{ {\scriptsize The observed and expected limit including 
                        systematic uncertainties as well as the theory cross section for 
                        $\sigma(p \bar p \to XX ) \times Br(\tilde g \tilde g \to 3 jets + 3 jets)$ 
                        where X = $\tilde g, \tilde q$, versus gluino invariant mass. The RPV gluino cross-section is from
                        PYTHIA and is corrected by an NLO k-factor.
 }}
    \label{figure8}
%    \end{minipage}
\end{center}
\end{figure}

\subsection{ Top + MET}

We conclude with a a search for a new particle T' 
decaying to top quark via T' $\to$  t + X, where X is an invisible particle\cite{ref:cdf-tprime-paper}. 
In a data sample with 4.8 fb$^{-1}$ of integrated luminosity collected by the CDF II detector, the search is conducted
for pair production of T' in the lepton+jets channel, $p \bar p \to  t \bar t + X + X \to  l\nu b q q\bar b + X + X$. 
Such process would produce extra missing energy and the key observable used in the analysis is the transverse mass 
distribution of the lepton-missing energy system, which in absence of new physics corresponds to the reconstructed
W transverse mass.
The results are primarily interpreted in terms of a model where T' are exotic fourth generation quarks 
and X are dark matter particles\cite{ref:feng-paper}. 
Current direct and indirect bounds on such exotic quarks restrict their masses to be between 300 and 600 GeV/c$^2$, 
the dark matter particle mass can be anywhere below $m_{T'}$. 
The data are consistent with standard model expectations, and CDF sets a 95\% confidence level limits 
on the generic production of TT' $\to  t \bar t$ + X + X, by performing a binned maximum-likelihood fit in 
the $m_W$ variable, allowing for systematic and statistical fluctuations via template morphing. 
The observed upper limits on the pair-production cross sections are converted to an exclusion 
curve in the mass parameter space for the dark matter model involving fourth generation quarks.
The current cross section limits on the generic decay, $T'\to  t+X$, may be applied to the many 
other models that predict the production of a heavy particle T' 
decaying to top quarks and invisible particles X, such as the supersymmetric process $\tilde t \to t + \chi^0$.
Applying these limits to the dark matter model CDF excludes 
fourth generation exotic quarks T' at 95\% confidence level up to m$_{T'}$ = 360 GeV/$c^2$ for m$_X <$  100 GeV/$c^2$ .

\begin{figure}[h]
\begin{center}
%  \begin{minipage}{2.5in}
\psfig{figure=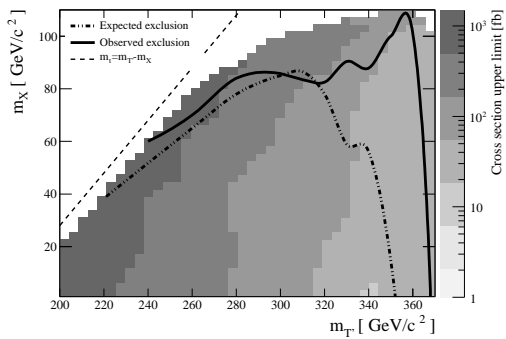,height=2.0in}
 \caption{ {\scriptsize Observed versus expected exclusion in (mT', mX ) along with the cross section upper limits.
}}
    \label{figure9}
%    \end{minipage}
\end{center}
\end{figure}

\section{ Conclusions}

The CDF and D0 experiments are actively collecting and analyzing data at the TeVatron collider.
New physics is searched in a broad manner, using different approaches.
In signature based analyses the data are scanned for anomalies 
pointing to indications of new physics, while many dedicated searches for specific models are pursued, using
the largest possible statistical samples.
New results on search for physics beyond the Standard Model are released almost daily. So far there is no
evidence for new physics and numerous limits on new particle masses and cross sections production are set.
A broader set of updated results can be found at:~http://www-d0.fnal.gov/Run2Physics/WWW/results/np.htm 
and~http://www-cdf.fnal.gov/physics/exotic/exotic.html.

\end{document}